\documentclass{article}
\usepackage{graphicx} 

  \usepackage{natbib}
\date{\today} 
\setcitestyle{round}
 
\title{Moderate Physical Perspectivalism} 
\date{\today}
\author{Emily Adlam  \thanks{Philosophy Department and Institute for Quantum Studies, Chapman University, Orange, CA92866, USA \texttt{eadlam90@gmail.com} }}

\begin{document}

\maketitle

\begin{abstract}  
Recent developments in foundations of physics have given rise to a class of views  suggesting that physically meaningful descriptions must always be relativized to a physical perspective. In this article I distinguish between strong physical perspectivalism, which maintains that all facts must be relativized to a perspective, and moderate physical perspectivalism, which maintains that all empirically meaningful descriptions must be relativized to a perspective.  I argue that both scientific evidence and philosophical considerations support moderate physical perspectivalism over strong physical perspectivalism. In particular, motivations connected to epistemic humility and the social nature of science are more compatible with the moderate approach. 
\end{abstract}

\section{Introduction}

Throughout the history of quantum mechanics it has been noted that observers seem to play a special role in the theory, and this has led to  speculations   that  there is something observer-dependent or perspectival about quantum mechanics. Separately, the central role of relational observables in General Relativity and quantum gravity also seems to point to a special role for internal perspectives in the formulation of a diffeomorphism-invariant physical theory. 

Such developments have motivated the development a class of views that I will refer to as  `physical perspectivalism.' These views often de-emphasize consciousness and instead  focus on   perspectives, defined by physical systems which play the role of a reference frame. A common thread running through this class of views is the idea that   it is a  mistake to try to formulate a `view from nowhere,' and this statement is often interpreted to mean that there cannot be \emph{any} kind of fact about physical reality which is not relativized to a perspective.

However, it is possible that the tendency to move immediately to this particularly strong form of perspectivalism has obscured some of the the insight that could be offered by the perspectival approach. We should also consider intermediate options, such as the view that I will call  `moderate physical perspectivalism,' which maintains that \emph{empirically meaningful} facts must be relativized to physical perspectives, but also admits the existence of some perspective-neutral facts.

    My goal in this article is, first,  to articulate more clearly   the distinctions   between  strong physical perspectivalism and moderate physical perspectivalism, and between epistemic perspectivalism and physical perspectivalism. And second,   to argue that both scientific evidence and philosophical considerations point more strongly towards moderate physical perspectivalism than strong physical perspectivalism. I will begin in section \ref{perspectives} by explaining the scientific case for physical perspectivalism and differentiating between strong and moderate physical perspectivalism. In section \ref{scientific} I will argue that the scientific evidence currently favours moderate physical perspectivalism over strong physical perspectivalism.  Then in section \ref{motivations} I will  explore the relationship between epistemic perspectivalism and physical perspectivalism,  arguing that the epistemic considerations which motivate epistemic perspectivalism may provide some motivation for moderate physical perspectivalism but they do not  provide any compelling motivation for strong physical persepctivalism.  Finally in section \ref{methodological} I will consider some methodological recommendations for scientific practice which might follow from moderate physical perspectivalism.

\section{Perspectives in Physics \label{perspectives}}

Recent developments in the foundations of physics have emphasized the role of perspectives in physical science. The motivations for this are twofold. First, in the context of quantum mechanics it has been argued that some form of perspectivalism may help to resolve the Wigner's Friend paradox \citep{Wigner1995} and various Extended Wigner's Friend paradoxes \citep{Bong_2020}. Second, in the context of General Relativity and quantum gravity it has been argued that physically meaningful observables  must be understood in relational terms,  meaning that they must be relativized to something like a reference frame or `perspective' \citep{1996cr}. The fact that both theories exhibit perspectival elements has been identified as a possible connection  which might help make progress towards a unified theory of quantum gravity \citep{vidotto2022relationalontologycontemporaryphysics}. This line of thought  goes right back to  \cite{bohratomtheorie}, who argued that there is no `view from nowhere' from which quantum systems can be described, and  connected this idea to relativistic principles: `\emph{the theory of relativity reminds us of the subjective character of all physical phenomena.}' Thus it seems potentially very important to understand exactly what form of perspectivalism is indicated by these developments.

In the Wigner's friend scenario \citep{Wigner}, an observer Alice performs a quantum measurement on a system $S$ inside a closed laboratory while another observer Bob looks on. Alice presumably now knows a definite value for some variable of $S$, but if Bob describes the whole scenario using standard unitary quantum mechanics, he predicts that Alice and $S$ are now in a state $\psi_{AS}$ corresponding to a superposition of all of the possible measurement outcomes, so to him it appears as if Alice has \emph{not} seen a definite value for any variable of $S$.  Thus this scenario motivates the adoption of some kind of perspectival view, in order that we can say that the state of $S$ is different relative to Alice and to Bob \citep{Dieks_2022}.

  Meanwhile, a central feature of General Relativity is the fact that it exhibits diffeomorphism-invariance \citep{Wallacenew,Earman2002-EARTMM} and many physicists expect that a successful theory of quantum gravity will also exhibit  diffeomorphism-invariance. In such a theory it is usually assumed that  histories related by a diffeomorphism are  one and the same history, so physically meaningful observables must be invariant under diffeomorphisms \citep{Wallacenew,Earman2002-EARTMM}. And it appears that the class of diffeomorphism-invariant observables which are best suited to express the kind of empirical content available to observers like us are  \emph{relational observables} \citep{pittphilsci4223}, e.g. observables of the form `the position of the pendulum relative to the clock reading a time $t$.' 

One influential approach to this issue, originated by \cite{Rovelli_2002}, makes use of the notion of a `partial observable,' i.e.  a physical quantity like position, mass and charge, `\emph{to which we can associate a measuring procedure leading to a number.}' Partial observables are not diffeomorphism-invariant and hence not physically real according to the standard interpretation of diffeomorphism-invariant theories, but the \emph{relations} between partial observables are diffeormorphism-invariant, and thus these relations constitute `complete observables' which are physically real according to the standard interpretation. Although  partial observables are not real on their own, we can arrive at a physically meaningful description in terms of partial observables by means of adopting a `perspective' to which these observables can be relativized.  Thus it is possible for us to `observe' partial observables even though they are not  individually physically real; we can observe them provided that they are relativized appropriately to our measuring instrument or body \citep{rovelli2022philosophical}. So this way of thinking about observables in a diffeomorphism-invariant theory leads naturally to a form of perspectivalism.

\subsection{ Physical Perspectivalism}

The increasing emphasis on perspectives   in the foundations of physics has given rise to a class of views which I will refer to as `physical perspectivalism,' suggesting that   physically meaningful descriptions must always be relativized to a perspective. 

It is important to emphasize that physical perspectivalism is not primarily about knowledge. This is clear from the role that physical perspectivalism is supposed to play in the Wigner's Friend experiment. In this experiment, if we think the state  $\psi_{AS}$ assigned by Bob merely reflects Bob's lack of knowledge about Alice's measurement outcome, we will naturally conclude that the true state of Alice and $S$ is $ \phi^{i}_{AS} = | i \rangle_A \otimes | i \rangle_S$, where $i$ corresponds to the definite value that Alice has actually seen.  But if unitary quantum mechanics is correct, it must be the case that when Bob subsequently performs measurements on Alice and $S$, he will always see outcomes consistent with the state $\psi_{AS}$, which are different from the outcomes we'd expect to see if the true state were either $\phi^{i}_{AS}$ or some probabilistic mixture of states of the form $\phi^{i}_{AS}$. Therefore the idea that the state is $\psi_{AS}$ relative to Bob cannot be merely a characterization of Bob's ignorance, but must be a substabtive assertion  about the actual dynamics that will take effect when Bob interacts with Alice and $S$. To resolve the puzzle we need  physical perspectivalism to be a claim about physics rather than  epistemology. 

 Since physical perspectivalism is not about knowledge, the `perspectives' to which facts are relativized   need not be associated with possible knowers, so consciousness need not play any special role in this kind of perspectivalism.
 Thus although some physical perspectivalists, such as  \cite{https://doi.org/10.48550/arxiv.2107.03513} and \cite{Cavalcanti_2021}, do focus   on perspectives associated with conscious agents, it's not clear that this limitation is well-motivated. Other physical perspectivalists have employed a more liberal notion of a perspective:  any physical system can act as the origin of a coordinate system relative to which we may describe the position of other physical systems, and physical perspectivalism generalizes this idea, suggesting that any physical physical system can anchor an `internal view' relative to which we may describe the  degrees of freedom of other physical systems.  

 One example of this approach  is Rovelli's Relational Quantum Mechanics \citep{1996cr}, which maintains that quantum descriptions must always be relativized to an `observer,' but then stipulates that any physical system can play the role of an observer. Another   is Dieks' perspectival quantum realism, which posits that `\emph{ objects may possess different, but equally objective properties with respect to different physically defined perspectives}' \citep{Dieks_2022}. Dieks emphasizes that `\emph{realism should not focus on agents and their subjective points of view; if perspectivalism is to be part of a realist scheme, the perspectives in question
should be defined with respect to physical systems.}' 

Physical perspectivalism also appears to be a motivator for  the quantum reference frame (QRF) formalism \citep{delahamette2021perspectiveneutral,article3,2020qctloe,2020acop,2020htsbrqc,giacomini2021quantum}, which provides a suite of tools for defining an internal view relativized to  a given physical system, and for switching from one such internal view to another. Proponents of this formalism talk about `\emph{jumping into the perspective of a quantum system}' \citep{2020acop} but this  is presumably not intended to suggest that quantum reference frames generally have \emph{conscious} perspectives. Rather the idea is that in a universe without a fixed background spacetime we must describe physics relative to a physical system, and if that system is quantum in nature then we will thereby be describing physics relative to a quantum reference frame.

\subsection{Strong Versus Moderate Physical Perspectivalism \label{strong}}

Let us differentiate two possible versions of physical perspectivalism. The first, which I will refer to as \textbf{strong physical perspectivalism}, maintains that \emph{all} facts about physical reality must be relativized to a (physical) perspective. The second, which I will refer to as \textbf{moderate physical perspectivalism}, maintains that \emph{empirically meaningful} facts about physical reality must be relativized to a (physical) perspective, where we say a fact is `empirically meaningful' if it gives a description of some part of reality in terms of phenomena of the kind that could be directly experienced by a realistic observer.  Unlike strong physical perspectivalism, moderate physical perspectivalism allows that there may exist some perspective-neutral facts about physical reality, but it maintains that such facts are typically quite abstract and  do not have direct empirical content.

Perspectivalists are fond of the mantra `there is no view from nowhere' \citep{pittphilsci19664,Berghofer2024-BERQRA-3}, and strong physical perspectivalists typically interpret this statement as asserting that there cannot exist physical facts of any kind which are not relativized to a perspective. However, moderate physical perspectivalists would interpret this statement differently.  One natural way to understand the term `view' is to think of it as referring to a description of some part of reality in terms that we ourselves can conceptualize or visualize - i.e. a description expressed in  empirically meaningful language referring to phenomena of the kind that could be directly observed by a realistic observer. And moderate physical perspectivalism does indeed maintain that any such `view' must be relativized to a physical perspective, so there is a sense in which it  upholds the idea that `there is no view from nowhere.'

Even without appealing to the scientific results discussed in section \ref{perspectives}, there are good reasons to believe that something like   moderate physical perspectivalism is true. 
For   although scientific theories are often formulated in an abstract, third-person way,   in order to extract from a theory  predictions which are useful for an embodied observer in a specific physical scenario we necessarily have to  move away from an abstract third-person description and adopt some kind of first-person view. As \cite{van2008scientific}  puts it, `\emph{If someone is to use (special relativity) to predict the behavior of electrically charged bodies in motion, bodies with which s/he is directly concerned, choice of a coordinate system correlated to a defined physical frame of reference is required. The user must leave the God-like reflections on the structure of spacetime behind in order to apply the implications of those reflections to his or her actual situation.}' That is, realistic observers necessarily have experiences from within a particular embodied perspective, so any empirical content of the kind that could be experienced by a realistic observer must at least implicitly appeal to the perspective of a possible observer who could have the corresponding experience.  

Moderate physical perspectivalism also seems like  a good way of capturing the features of quantum mechanics and relativity discussed in section \ref{perspectives}. We have seen that the Wigner's friend paradox pushes us towards  relativizing quantum states  to perspectives, and quantum states encode predictions for measurement outcomes that could be obtained by observers,  so they   fall into the category of `empirically meaningful descriptions.' Thus the relativization of states to perspectives in a quantum context looks like an instance of the idea  that empirically meaningful descriptions must be relativized to a perspective. Meanwhile, in the partial/complete observable framework, statements about phenomena that could be experienced by  observers  are necessarily expressed in terms of partial observables, and the formalism tells us that partial observables must always be relativized to some physical system or `perspective,' so again we have an instance of the idea that empirically meaningful facts should be relativized to a perspective. 

Additionally, since perspectivalism has been cited as a potential route to unifying quantum mechanics and relativity, we should consider what is suggested by the combination of both theories. At present it is not straightforward to extract empirical content from existing proposals for full theories of quantum gravity, so I will instead address this question by considering some general consequences of the  Dirac quantization procedure \citep{dirac1981principles}, which is the usual approach to  quantizing theories subject to constraints. Since many physicists expect that a theory of quantum gravity will satisfy diffeomorphism invariance just as General Relativity does, a natural way of approaching the problem is to use   Dirac quantization with a diffeomorphism constraint. 

In the process of Dirac quantization, we start from a `kinematical' Hilbert space encoding the most general possibilities, and   we impose constraints in order to move to a `physical' Hilbert space  encoding only states that are physically possible,  in the sense that they respect the symmetries associated with the constraints. But in the case of a diffeomorphism constraint the physical Hilbert space is quite abstract and contains redundancies in its representations, so `\emph{in order
to make operational sense out of physical phenomena, we must make additional choices to fix
these redundancies}' which can be done by \emph{choosing
a system from the perspective-neutral picture to
serve as our reference frame}' \citep{2020acop}. That is, once again we find that in order to extract empirically meaningful content it is necessary to adopt an internal reference frame; so current ideas about quantum gravity also seem to support the idea that empirically meaningful descriptions must be relativized to perspectives.

\section{Scientific Evidence \label{scientific}}

It appears that considerations from quantum mechanics and relativity do provide support for the idea that empirically meaningful descriptions typically have to be relativized to a perspective. But this in and of itself does not tell us whether we should adopt moderate or strong physical perspectivalism. To make that choice, we must decide whether these considerations suggest that  \emph{all} facts about physical reality must be relativized to a perspective, or whether they instead support the existence of  some perspective-neutral facts about physical reality, over and above descriptions  relativized to perspectives.

First of all, consider the quantum case. Accepting for the moment that   quantum states must be relativized to physical perspectives, does that suggest that \emph{all} physical facts must be relativized to physical perspectives? One way to argue for such a  conclusion   might involve maintaining that  that quantum mechanics is complete, i.e. that the quantum state of a system is a complete description of all of its physical properties \citep{Maudlin1995-MAUTMP}. In that case, it would follow from the  relativization of quantum states that  there cannot be any properties  of  physical systems which are not relativized to a perspective.   

However,  there still appear to be certain kinds of \emph{relations} which must remain perspective-neutral. For example, the  only reason why Wigner's Friend scenario poses a problem in the first place is because we are assuming that both Alice's observations and Bob's observations should be in accordance with the predictions of quantum mechanics. That is, the formulation of the paradox relies on the assumption that there exists a certain kind of relation between these perspectives - although Alice and Bob may differ on the outcomes of specific measurements, both perspectives exhibit the same kinds of regularities. This, if true, is a fact  about physical reality. Moreover, the description of the paradox assumes that this is a  perspective-neutral fact,  since we cannot arrive at any paradox or contradiction here if we are not able to compare the contents of the two perspectives from a third-person point of view.   So it seems impossible to even formulate the motivation for physical perspectivalism in quantum mechanics without a commitment to some kind of perspective-neutral facts about the relations between perspectives, and thus the quantum considerations described in section \ref{perspectives} appear more compatible with moderate physical perspectivalism  than strong physical perspectivalism.

Second, consider the relativistic case. We have seen that within the partial/complete framework, empirical content is encoded in the relativized partial observables; but of course this  framework also posits complete observables in addition to partial observables.  Complete observables are diffeomorphism-invariant, delocalised, and do not undergo change over time, so they do not appear to be relativized to a perspective; facts about complete observables are perspective-neutral in the strongest possible sense. And although these facts do not   directly describe possible experiences that any observer could have,  they are clearly still facts about physical reality, since the standard interpretation of diffeomorphism-invariant theories tells us that \emph{only}  the complete observables are physically real. 

More generally, it is common to formulate diffeomorphism-invariant theories such as General Relativity  in terms of tensors, because tensorial descriptions `\emph{encode the physics as
experienced in any local spacetime reference frame at once: an abstract tensor has to be contracted with a
vector frame, in order to determine the numbers that a corresponding observer would find in a measurement
of the quantities embodied by the tensor. In this sense, tensors abstractly constitute a description of the
local physics before a choice of reference frame has been made; they are reference frame perspective-neutral
structures}' \citep{delahamette2021perspectiveneutral}. That is, the purpose of writing the equations of the theory in tensorial form is precisely to  capture the perspective-neutral content of the theory. And again, these tensorial descriptions are clearly intended to encode facts about physical reality, since they are considered an equally valid formulation of the theory, and indeed many people working in the foundations of physics hold the view that all spacetime theories should be formulated in a coordinate-free way based on objects like tensors \citep{malament2012topics,0d31a2be-2c1a-3be3-852e-e5c70359dae5,Anderson1971-ANDCIA}.  So our current understanding of diffeomorphism-invariant theories  appears to presuppose the existence of perspective-neutral facts, and thus  relativistic considerations  appear more compatible with moderate physical perspectivalism rather than strong physical perspectivalism.

Finally, with regard to quantum gravity, we have seen that in the context of Dirac quantization it is necessary to use something like the QRF formalism to extract empirical content out of a state defined on the physical Hilbert space. However, it remains the case that `\emph{the physical Hilbert space encodes and links all internal frame perspectives, which is why it is also called the perspective-neutral
Hilbert space}' \citep{delahamette2021perspectiveneutral}. Thus  the global state $\Psi$ defined on the physical  Hilbert space represents `\emph{a perspective-neutral super structure that encodes, so to speak, all perspectives at once and requires additional choices to ‘jump’ into the perspective of a specific frame}' \citep{2020acop}. As argued by \cite{2020acop}, this perspective-neutral structure plays a similar role to Minkowski spacetime in special relativity, unifying all of the individual reference frames; it achieves this in virtue of  a quantum coordinate map which maps from the state $\Psi$ into the perspective of an individual quantum reference system, so we can 
always  transform between internal reference frames by composing an inverse quantum coordinate map from the first reference frame  to the global state $\Psi$ with another quantum coordinate map from  $\Psi$ to the second reference frame \citep{delahamette2021perspectiveneutral}. So although the Dirac quantization procedure does support the idea that empirically meaningful facts must be relativized to perspectives, it  also suggests the existence of a perspective-neutral background structure encoding the relationships between all the various physical perspectives, and therefore it provides evidence for the  existence of perspective-neutral facts in a quantum theory of gravity.   Thus   scientific considerations based on quantum mechanics, relativity and their combination all seem to point towards moderate rather than strong physical perspectivalism.

 \section{Philosophical Intuitions\label{motivations}}

The recent perspectival turn    in the foundations of physics   has been dominated by what appears to be strong physical perspectivalism. There is some uncertainty here, because  the existing literature does not distinguish clearly between what I have called strong and weak physical perspectivalism, so it is possible that some  authors who appear to be advocating strong physical perspectivalism actually intend to be advocating something more similar to moderate physical perspectivalism. However, there certainly seem to be some inclinations towards a fairly radical form of perspectivalism - for example, \cite{brukner2015quantum} takes Wigner's Friend scenarios to show that `\emph{ ``facts" can only exist relative to the observer}' and \cite{Cavalcanti_2021} suggests that these  scenarios may motivate a move towards a pragmatic rather than a correspondence theory of truth.

Since it appears that the scientific evidence does not clearly support strong physical perspectivalism,  this enthusiasm for strong perspectivalism is presumably  driven to some degree by philosophical convictions rather than just scientific evidence. Thus  in this section I will consider some of the philosophical intuitions lying behind   physical perspectivalism, and discuss whether they in fact favour strong or moderate physical perspectivalism.

\subsection{Rejecting the Cartesian-Hegelian Ideal \label{intermediate}}

One important motivation for   physical perspectivalism is the idea that modern physics has revealed the bankruptcy of what one  might call the traditional `Cartesian-Hegelian' metaphysics, in which it is assumed that the most fundamental description of reality is completely observer-independent and can be characterized in  wholly third-person terms. For example, \cite{cuffaro2021open} write `\emph{the ideal of an observer-independent reality is not methodologically necessary for science and ...
modern physics (especially, but not only, quantum theory) has taught us ... that there is a limit to the usefulness of pursuing this ideal.}’ 

Now, we have seen that both quantum and relativistic considerations  provide good reasons to question the Cartesian-Hegelian metaphysics.  However,   the Cartesian-Hegelian picture is  characterized not only by the idea that there \emph{exists} a  perspective-neutral description of reality, but by the stronger claim that this perspective-neutral description constitutes the most fundamental and complete characterization of reality. One way of rejecting the Cartesian-Hegelian picture is to deny the   \emph{existence claim}, which leads to strong physical perspectivalism; but another way of rejecting the Cartesian-Hegelian picture is to simply deny the claim about the   \emph{fundamentality} of the third-person perspective while leaving the existence claim alone, which  leads instead to something like   moderate physical perspectivalism, in which we stipulate that both perspectival and perspective-neutral facts exist and both are equally fundamental. That is, moderate physical perspectivalism suggests that although we may be able to map between the perspectival and perspective-neutral descriptions, first-person perspectives are an essential feature of reality as we experience it and we should not be seeking  to  eliminate them  
  in favour of a third-person view. Thus although  it may well be true that  quantum mechanics and/or relativity give us reasons to reject the traditional  Cartesian-Hegelian metaphysical picture, this does not mean we must be strong physical perspectivalists; moderate physical perspectivalism offers an alternative way of responding to this situation.

For example, given the central role of partial observables in our empirical experience, there is a sense in which the partial observables must be taken seriously as elements of reality even though they are not individually real according to the standard interpretation: as \cite{pittphilsci4223} puts it, `\emph{both spaces—the space of genuine (complete) observables and partial observables—are invested with physicality by Rovelli.}' Insofar as this is a correct characterization of Rovelli's views, one way to understand it would be to see Rovelli as investing the partial observables with physicality because he sees the  first-person perspective (relative to which partial observables are defined) as having equal validity to the third-person perspective (in which  complete observables are defined) - so this way of thinking about the relation between partial and complete observables does indeed look like a form of moderate physical perspectivalism.

\subsection{Epistemic Perspectivalism}

Other  philosophical intuitions lying behind the adoption of strong physical perspectivalism appear to be linked to previous philosophical perspectivalisms. There is a long    tradition of such philosophical views, stretching back at least to the work of   \cite{LeibnizManuscript-LEIM-3}, and continuing through the work of \cite{Nietzsche2007-NIEOTG-2},  \cite{Kuhn},  \cite{rorty1991objectivity}, and  \cite{Giere2006-GIESP} among many others. In general, such views maintain that all  knowledge is necessarily relativized to the perspective of the knower, and no matter how hard we try to achieve `objectivity,' we will never wholly escape our perspective to achieve perspective-neutral knowledge. As \cite{rorty1991objectivity} puts it, `\emph{the image of climbing out of our minds – to something external from which we can turn and look at them – needs to be replaced.}'  

Notably, these previous philosophical perspectivalisms have typically fallen into the category of what I will refer to as \emph{epistemic} perspectivalism - that is, positions which pertain to   what it is possible for beings like us to know or believe, rather than what exists. That is, when epistemic perspectivalists assert that `there is no view from nowhere,' this is not a claim about the existence or otherwise of perspective-neutral facts; rather it is asserting that the `view from nowhere' is not a perspective that can be taken up by any realistic observer. Relatedly,    since epistemic perspectivalism is  a claim about what certain kinds of beings can know or how they can think, it typically   concerned only with   `perspectives' associated with conscious subjects.  

Evidently then these previous philosophical perspectivalisms are different from physical perspectivalism. It does seem plausible that physical perspectivalism entails some form of epistemic perspectivalism; for if all facts, or all empirically meaningful facts, must be relativized to a perspective, then it seems natural to think that knowledge of such facts must also be associated with a perspective. But the converse implication is not so obvious. Many presentations of epistemic perspectivalism explicitly endorse the existence of an external perspective-neutral reality while maintaining we cannot have perspective-independent knowledge of it: for example, \cite{Massimi2018-MASFKO-4} defines perspectival realism as the view that `\emph{states of affairs about the world are perspective-independent; whereas our scientific knowledge claims about these states of affairs are perspective-dependent.}' 

 It is true that there exist some philosophical views which  would have the effect of erasing the distinction between epistemic and physical perspectivalism. For example, if one is committed to some form of ontological idealism  view which posits that (human) thought is the foundation of reality \citep{sep-idealism}, then the claim that agents cannot have knowledge which is not relativized to their own perspective leads inevitably to the claim that all physical facts must be relativized to a perspective. However,  such idealism  does not combine well with versions of physical perspectivalism which allow that all   physical systems can play the role of  perspectives - for a description relative to a non-conscious physical system is not known or perceived by anyone, so it would not make sense for a typical idealist to see this perspective as physically meaningful. Thus although it may be the case that the physical perspectivalists who focus on conscious perspectives are ultimately peddling a new form of idealism,  other physical perspectivalists who employ a more liberal notion of `perspective' cannot be considered idealists,   so in general physical perspectivalism must  be understood as a distinct position from epistemic perspectivalism. 

But nonetheless, many physical perspectivalists appear to be motivated by intuitions linked to epistemic perspectivalism.  In particular, one common  motivation for epistemic perspectivalism   is a kind of epistemic humility, based on the observation that all knowledge is possessed by physically embodied observers who are subject to various kinds of limitations in terms of their epistemic access to reality.   For example, Giere's scientific perspectivism \citep{Giere2006-GIESP} is motivated by consideration of   epistemic limitations to which we are subject, such as the limitations of our scientific instruments: `\emph{instruments are sensitive only to a particular kind of input. They are, so to speak,
blind to everything else. Second, no instrument is perfectly transparent. That is, the output
is a function of both the input and the internal constitution of the instrument}.' Giere argues that our instruments are perspectival because of the way in which existing theory influences their design and intended applications, and thus the scientific knowledge we obtain from them is also necessarily perspectival.

And some proponents of physical perspectivalism also seem to be motivated by a similar kind of epistemic humility. For example, \cite{rovelli2024princetonseminarsphysicsphilosophy} first notes that there are inevitable limitations on human knowledge -   `\emph{We live in this space between ignorance and certainty. The two extremes are of no interest. What matters to us is the space in between}' -  and then uses this observation to motivate  the conclusion that `\emph{reality is more tenuous than the clear-cut one imagined by the old physics models; it is made up of happenings, discontinuous events, without permanence, located with respect to one another and only existing relatively to one another.}' Here, epistemic considerations are used to motivate what appears to be a form of \emph{physical} perspectivalism. Similarly, \cite{pittphilsci16956} invokes Giere's arguments about the limitations of scientific instruments   to motivate a version of \emph{physical} perspectivalism  intended to address the Wigner's Friend paradoxes. It appears that these authors believe that the considerations which support epistemic perspectivalism also lend some support to physical perspectivalism. 

  And in fact, it seems plausible that epistemic humility provides some support for moderate  physical perspectivalism. Such an argument might go something like this: 

\begin{enumerate} 

\item Realistic observers are always subject to limitations in their epistemic access to reality. 

\item Therefore all of the direct empirical content that can be experienced by a realistic observer is filtered through the lens of   their epistemic limitations. 

\item Perspective-neutral facts about reality are, by definition, not filtered through the lens of any epistemic limitations. 

\item Therefore  perspective-neutral facts cannot express direct empirical content of the kind that could be experienced by a realistic  observer.

\end{enumerate}

 But can epistemic humility support \emph{strong} physical perspectivalism? Such an argument would presumably   go something like this: 

\begin{enumerate} 

\item Realistic observers are always subject to limitations in their epistemic access to reality. 

\item Therefore scientific enquiry  can only yield knowledge which is relativized to a perspective.

\item Therefore scientific enquiry cannot give us any evidence for the existence of facts which are not relativized to a perspective. 

\item Thus, applying Ockham's razor, we should believe that there do not exist any facts which are not relativized to a perspective.

\end{enumerate}

However, there are two important problems with this argument\footnote{There are other problems as well, but I will focus here on the two which I find most relevant to understanding the connection between epistemic and physcial perspectivalism.}. I will go through these two problems in some detail, because I think this will help to clarify the connection between epistemic and physical perspectivalism, and will therefore provide a clearer understanding of the philosophical motivations for strong physical perspectivalism.

\subsubsection{Self-Undermining}

The first problem with the argument is that its conclusion is inconsistent with its starting premise. For if  I accept that strong physical perspectivalism is correct, then I cannot meaningfully articulate the thought that I have limitations in my epistemic access to reality, and therefore I will not be able to affirm premise 1) in the argument above. 

To see this, consider the following claim, $C$: `There are features of reality which my perspective fails to grasp.' Claim $C$ is, if true, a fact about physical reality, and therefore the strong physical perspectivalist is obliged to maintain that $C$ must be relativized to a perspective. Yet which perspective could it possibly be relativized to? 

Claim $C$ cannot be relativized to my own perspective, since it explicitly makes assertions about features of reality which are external to my perspective. Could  claim $C$ be relativized to the perspective of some other system $X$? It seems not, because   in order to make such a statement from within the perspective of $X$, it would be necessary for the perspective of $X$ to have access to facts about the content of my perspective. But if the perspective of $X$ has access to such facts, that means there exist some systematic connections  between my perspective and the perspective of $X$, and the existence of those connections would be a fact about reality which is not relativized to any perspective, in violation of the central tenet of strong physical perspectivalism. We could try to say that the existence of such connections are themselves relativized to a third perspective $Y$,  but that does not seem to help us: for if the connections only exist within the perspective of $Y$, then it will still not be possible to assert claim $C$ relative to the perspective of $X$, since it is not a fact relative to $X$ that $X$ has access to facts about my perspective.  So no matter how far we push the regress of relativization, there does not seem to be any meaningful way we can assert claim $C$ in the context of strong perspectivalism.

One might think that although as a strong physical perspectivalist I cannot directly articulate the idea that I am limited in my access to the world, I might be able to observe   that \emph{other} agents are limited in their access to the world, and then appeal to some kind of symmetry principle in order to conclude that I myself must be similarly limited in my access to the world. But this kind of reasoning is impossible for a strong physical perspectivalist. For the invocation of a symmetry principle in this connection makes sense only if I regard the perspectives of others as equally real and meaningful as my own, and yet if I do that I am asserting some kind of perspective-neutral fact about the existence or meaningfulness of those other perspectives. As soon as I take the perspectives of other observers seriously enough to admit facts about their perspectives as meaningful constraints on my own knowledge, I am necessarily stepping outside of the bounds of strong physical perspectivalism.

Moroever, the issue here is not just about the coherence of a specific argument for strong physical perspectivalism: the problem is that strong physical perspectivalists are shutting themselves off from a  suite of very valuable scientific tools and insights. It is \emph{important} to recognise  that embodied observers are subject to specific epistemic limitations, in order that we can endeavour to understand those limitations;   for once we realise that some feature of our experience is actually an artefact of our own limitations, we can often get a better grasp of the nature of the reality lying behind those experiences.    For example, one way of thinking about the science of thermodynamics is that it is a resource-relative theory i.e. it describes what can be achieved by agents with certain fixed resources to manipulate systems,  and thus thermodynamics is a science which helps us understand the consequences of our specific epistemic and physical limitations \citep{Robertson_Prunkl_2023}. This way of thinking about scientific theories seems to be ruled out if we adopt a view which cannot acknowledge the potential existence of features of reality beyond the perspective of an individual agent.

So in fact, although epistemic humility is certainly laudable, it actually seems that   strong physical perspectivalism  prevents us from doing justice to the insights that follow from epistemic humility. If we want to  make use of such insights  we are better off adopting something like moderate physical perspectivalism, which allows us to maintain  the existence of a third-person reality within which our own perspective can be situated and thus allows us to coherently discuss the limitations of our own perspective. In so doing we of course acknowledge that we will never know all of the facts about  that third-person reality, but even simply acknowledging the ways in which our perspectives are limited can be a useful aid to progress. As \cite{Nagel1986-NAGTVF} puts it, `\emph{it is necessary to combine the recognition of our contingency, our finitude, and our containment in the world with an ambition of transcendence, however limited may be our success in achieving it. The right attitude in philosophy is to accept aims that we can achieve only fractionally and imperfectly, and cannot be sure of achieving even to that extent.}'

\subsubsection{Moderate epistemic perspectivalism}

The second problem with this argument pertains to the move from premise 1) to premise 2). For  the  fact that observers are always subject to limitations in their epistemic access to reality does not imply  that \emph{all} knowledge is relativized to an individual perspective, unless we make the additional assumption that there is no   possible way in which individual observers can transcend their epistemic limitations, and this assumption seems questionable. 

For example, consider the following statement of perspectivalism:  `\emph{(various different forms of perspectivalism) share the general idea that there is no “view from nowhere”, and that scientific knowledge cannot transcend a human perspective}' \citep{Ruyant2020-RUYPRA}. There is an essential ambiguity in this kind of statement. If the claim is that scientific knowledge cannot transcend the perspective of an \emph{individual human}, then the claim is wrong. Scientific knowledge is created and possessed by an entire epistemic community, and in that sense it does transcend the perspective of any one individual. Whereas if the claim is that scientific knowledge cannot transcend certain limitations that are shared by all humans, then the claim is probably right, but it's no longer clear that this claim entails that there is no `view from nowhere,' in the strongest sense of that phrase. For knowledge  belonging to our entire epistemic community is not a view from any specific place and time; it is delocalized over many individual perspectives at different places and times, and thus even if it is not precisely a view from \emph{nowhere}, it is nonetheless also not a view from any particular location.

Moreover, the knowledge of the community is not just the sum of knowledge from a variety of individual perspectives: it also includes the \emph{relations} between these perspectives, and a fact about the relation between two perspectives is  not itself directly accessible within either of those perspectives.  Thus by acting as a community we can broaden our perspective away from the view of any individual, so this methodology represents progress towards knowledge of facts which are not relativized to a perspective. For example,  \cite{Nietzsche2007-NIEOTG-2} espoused a version of perspectivalism in which, by acting as a community, we may  `\emph{approach “objectivity” (in a revised conception) asymptotically, by exploiting the difference between one perspective and another, using each to overcome the limitations of others}' \citep{sep-nietzsche}. More recently, \cite{longino1990science} argues that the process of discussion and critique of scientific ideas across the community is a defining property of scientific practice which helps to  `average out' individual limitations and biases in order to produce knowledge which has more weight than the conclusions that could be drawn by any individual alone.   Of course, these social practices are not going to transcend all of our limitations, but nonetheless they represent progress towards knowledge of something perspective-neutral. 

 Moreover, if we look at the reality of actual scientific practice it is clear that science \emph{does} and always has included both first-person and third-person views on the world. This is particularly clear  in the theory of special relativity, which provides us with tools for describing physics relative to individual perspectives, but \emph{also} provides us with Minkowski spacetime  in order to formulate a third-person, `gods-eye' description of the way in which all of the perspectives are related \citep{van2008scientific}. It would  be very strange to argue that either the first or the third-person descriptions should be eliminated from special relativity: we need  first-person descriptions relative to reference frames in order to obtain useful predictions from the theory, but we also need the third-person descriptions to reveal the underlying  invariant structures. Although authors may disagree on whether the reference frame description or the Minkowski spacetime description is more `fundamental,' there is no reason to think that either kind of description is illegitimate or meaningless.

These considerations motivate what we might call `moderate epistemic perspectivalism,' - a class of views which emphasize the essential role of first-person   knowledge in science, but which nonetheless maintains that it is sometimes possible to transcend the limitations of one's individual perspective.  For example, \cite{shimonyreality} argues that the  epistemology of science  can be understood as a process of `closing the circle,' which  `\emph{envisages the identification of the knowing subject (or more generally, the experiencing subject) with a natural system that interacts with other natural systems. In other words, the program regards the first person and an appropriate third person as the same entity.}' The methodology of closing the circle does not involve eliminating either the  internal or the external view in favour of the other;  it is simply a process of learning to map between them.   Similarly, \cite{Nagel1986-NAGTVF}   emphasizes that in science and in intellectual activity more generally we are constantly navigating between first-person and third-person views or reality, and he contends   that we should resist the tendency to eliminate one type of view in favour of the other: `\emph{the correct course is not to assign victory to either standpoint but to hold the opposition clearly in one's mind without suppressing either element.'}  Thus both Nagel and Shimony appear to be advocating what I have called moderate epistemic perspectivalism.

And the possibility of some kind of moderate epistemic perspectivalism  is highly relevant to physical perspectivalism. For if it turns out that the relevant epistemic considerations support only moderate epistemic perspectivalism and not strong epistemic perspectivalism, then surely insofar as those epistemic considerations imply anything about physical reality, they can only support moderate physical perspectivalism and not strong physical perspectivalism. After all, if we conclude that in fact observers in our actual world \emph{do} sometimes have  knowledge about features of reality which are not relativized to an individual perspective, it is then surely impossible to maintain that there do not \emph{exist} any features of reality which are not relativized to an individual perspective, and thus strong physical perspectivalism collapses. Thus it seems to me that the argument from epistemic humility is not in fact a good motivation for strong physical perspectivalism; these epistemic considerations point much more strongly towards moderate physical perspectivalism.

\subsection{First-person plural views \label{communities}}

In light of  concerns about the social nature of science, a number of epistemic perspectivalists have adopted what we might call a first-person-\emph{plural} view, where perspectives are associated not merely with single individuals but also with entire epistemic communities. Thus epistemic perspectivalists can accept that indeed, social features of science allow us to transcend the perspectives of individual observers, but still maintain that the resulting knowledge is  relativized to a perspective - it is simply the perspective of the whole epistemic community, rather than an individual observer. For example, \cite{rorty1991objectivity} argues that `\emph{Whatever good the ideas of `objectivity' and `transcendence' have done for our culture can be attained equally well by the idea of a community which strives after both intersubjective agreement and novelty.}' 

And in light of the concerns expressed in the previous section, physical perspectivalists might be tempted to make a similar move. This would involve arguing that in addition to facts relativized to the perspectives of individual observers, there also exist facts relativized to some kind of emergent perspective that results from information being shared across a community, so physical perspectivalists can acknowledge the role of the community in the the acquisition of scientific knowledge.  

One example of such a view is Healey's `desert pragmatism' \citep{ Healey2012-HEAQTA}  in which quantum states and measurement  outcomes are relativized to the situation defined by a `decoherence environment,' i.e. a region of spacetime in which environmental decoherence stabilizes the value of some variable such that it can meaningfully be assigned a value. Since decoherence will spread rapidly through a community of macroscopic observers, it is reasonable to expect that the entire human community belongs to the same decoherence environment, so Healey's picture suggests that our entire epistemic community shares a set of  relativized facts about quantum states and outcomes.  

Similarly, \cite{pittphilsci16956} puts forward a  notion of `perspectival objectivity,' in which  `\emph{a scenario in which some feature of the world is in part a function of the
agent perspective while at the same time, given such a perspective that is inescapably shared
between similar agents, there is an (intersubjectively) objective fact of the matter concerning
that feature.}' \cite{pittphilsci16956} appears to be arguing  that this kind of objectivity can exist even in the context of physical perspectivalism,   since he  explains the motivation for the view as follows: `\emph{to attempt to accommodate the recent claims from quantum foundations that quantum mechanics rules out the possibility of “observer-independent facts”.}'  That is,  Evans is positing that  in the context of physical perspectivalism, the process of sharing information can bring into being a higher-level emergent perspective associated with a whole epistemic community, and thus  we can still do justice to the important role of intersubjective sharing of information as a means of achieving a higher level of objectivity.

 Now, both of these approaches work well in the context of moderate physical perspectivalism. For example, Evans' notion of perspectival objectivity would be a helpful way for the moderate physical perspectivalist to explain how empirically meaningful facts relativized to individual perspectives can  nonetheless be understood as having a kind of objectivity in virtue of  the way in which they are shared  between the perspectives of different agents.  

But can such approaches be employed in the context of \emph{strong} physical perspectivalism? This seems challenging, since   strong physical perspectivalism is committed to denying the existence of any mechanisms which could bring about connections between the perspectives of different agents. Moderate physical perspectivalists can maintain that there exists some kind of perspective-neutral  underlying  structure which connects all of the perspectives together and explains the relations between them, such as Minkowski space or the global state $\Psi$ described in section \ref{scientific}, but strong physical perspectivalists cannot allow such a possibility: they would have to insist that facts about such a  structure must themselves  be relativized to something, and thus, as \cite{Riedel_2024} puts it, we inevitably end up with an `\emph{iteration of relativization.}'   So   we can't simply get rid of external reality and yet maintain that information is really shared across a community; the   first-person plural approach in the context of strong physical perspectivalism amounts to  telling the same  story as in the  moderate case  but then  pulling out the rug in a way that destabilizes the whole view. 

We can see this in action with regard to Evans'  concept of perspectival objectivity. For in order for this concept to apply to some set of observers $\{ O \}$ in certain specific circumstances, there must exist certain facts about a)  the nature of the observers in $\{ O \}$, b)   their current circumstances, c)  the ways in which the observers are constrained in their access to the world, and d) the relations that hold between different observers, such as the fact that intersubjective sharing of certain kinds of information is possible for them.  But in the context of strong physical perspectivalism we are not allowed to postulate any facts which are not relativized to a perspective, so which perspective could facts a) - d)  be relativized to? 

We might try to say that these facts are simply relativized to  some particular observer $O'$ in the set $\{ O \}$. However,  in that case it is true only relative to that  $O'$ that information is shared across the set of observers, so  the  `shared' information is  no more objective than any other piece of information relativized to $O'$: the higher level of objectivity which is supposed to be conferred by the fact that the information is shared by the whole community is lost. 

We might try to say that the facts a) - d) are true relative to the  joint perspective of all of the observers in the set $\{ O \}$. But the joint perspective of a community does not simply come into being as a fundamental unit; rather the community is \emph{brought} into being by dynamical interactions which create correlations between individual perspectives, and in order to make sense of this process we need to be able to describe these dynamical interactions in a way which does not presuppose the existence of the joint perspective which is supposed to be created by those very interactions. In addition,   if we simply postulate the whole epistemic community as a fundamental unit,   we will be left with no principled way to distinguish between groups of observers who share a joint perspective and groups who  do not: it is precisely the existence of perspective-neutral facts about the spread of information which allow us to identify when an epistemic community has been formed. Thus in order to do justice to the dynamical, participatory formation of an epistemic community, we must be able to offer a perspective-neutral description of the way in which  dynamical processes create connections between perspectives.

Finally, we might try to say that the facts a) - d) are true  relative to some other observer outside of the set $\{ O \}$.   But it is hard to see how this would yield any meaningful notion of objectivity. What Evans wants is for the members of the community $\{ O  \}$ to be able to see their \emph{own} scientific judgements as  (quasi) objective, in virtue of information being shared across the whole community;  and it is no help from the point of view of an observer in the set $\{ O\}$ that some \emph{other} observer thinks the information is shared across the whole community. 

Moreover, the need for perspectival objectivity to be backed up by   perspective-neutral facts is clear from Evans' motivating examples. Evans cites colour vision and causation as  perspectivally objective phenomena, but the perspectival objectivity of these phenomena are grounded on underlying perspective-neutral facts about how our individual perspectival experiences of these phenomena are related. For example,  Evans argues that although colour is not part of the fundamental furniture of the world, it has a kind of `objectivity' in virtue of the fact that most members of our epistemic community agree about judgements of colour. But that agreement does not exist in a vacuum. The members of our epistemic community agree about colour because there exists an underlying mechanism according to which objects reflect certain  frequencies of light and the reflected light causes observers  like us to have certain kinds of experiences. And the facts about which frequencies of light are reflected by a certain object are not relativized to any perspective, which is precisely what allows these facts to ground  connections between colour experiences featuring in two different perspectives. If there were no such mechanism connecting our experiences of colour vision, we \emph{wouldn't} in general agree about judgements of colour, and thus there would be no perspectival objectivity. 

For these reasons it seems to me that we cannot invoke anything like Evans' notion of perspectival objectivity or Healey's desert pragmatism in the context of strong physical persectivalism. These approaches work much better if we think of them as forms of moderate physical perspectivalism. So there exist a number of interesting routes to explore here for the moderate physical perspectivalist, but options appear very limited for the strong physical perspectivalist.

\section{Methodological Recommendations \label{methodological}}

A final important philosophical motivation for strong physical perspectivalism is the idea that our difficulties in interpreting quantum mechanics arise from our unwillingness to put aside the na\"{i}ve classical notion of an observer-independent universe.

However,  in fact moderate physical perspectivalism can  offer  similar kinds of resolutions to these issues. For  order to diagnose the interpretive difficulties of quantum mechanics as arising from neglect of the first-person perspective, it is  not necessary to reject \emph{all} perspective-neutral facts: the problem may not be that we are trying to  achieve  objectivity, but rather that we are doing it too much, or in the wrong way.

For example,  \cite{Nagel1986-NAGTVF} offers some methodological recommendations about how to properly balance and first and third-person views on reality:  we should avoid `\emph{excessive impersonality,}' and we should avoid `\emph{false objectification.}' Excessive impersonality refers to the tendency to become so focused on the objective standpoint that we forget we are also `\emph{creature(s) with an empirical perspective and individual life.}' False objectification refers to cases in which `\emph{the success of a particular form of objectivity in expanding our grasp of some aspects of reality (tempts) us to apply the same methods in areas where they will not work.}'  

And indeed, it appears that in the context of   moderate perspectivalism these recommendations could plausibly help us with our quantum woes. For example, one might argue that `excessive impersonality' is committed by the proponents of the Everett interpretation \citep{manyworlds}. For quantum mechanics is, as proponents of perspectival approaches are fond of pointing out, a formalism developed by human observers  as a characterisation of specific aspects of the world accessible to creatures like us. But the Everett approach takes this formalism and takes it  literally as an objective, observer-independent characterisation of the world as a whole. From the point of view of a moderate physical perspectivalist, it may appear  that the Everettian move amounts to disregarding the pragmatic, human-oriented origins of the formalism and turning it into an impersonal description in a way that is not justified by the empirical facts. 

Similarly, one might argue that `false objectification' is committed by the proponents of wavefunction collapse views \citep{GRW,sep-qm-collapse}. In earlier scientific theories we have achieved significant success by taking wavelike structures seriously as literal elements of reality - as for example in our understanding of sound and in the pre-quantum formulation of electromagnetism. Wavefunction collapse views extrapolate this to quantum mechanics and seek to take the quantum wavefunction literally as an element of reality similar to a sound wave or electromagnetic wave, but from the point of view of a moderate physical perspectvalist, it may appear that this is a case of us being tempted to apply a particular form of objectification in an area where it will not work. 

 These examples indicate that although our difficulties with quantum mechanics may indeed have something to do with failures to appreciate the important role of perspectives in physics, it is not necessary to go all the way to strong physical perspectivalism to rectify these failures. Moderate physical perspectivalism already offers important new ideas about how to think about quantum descriptions, and does so without encountering the severe conceptual difficulties that affect approaches based on strong physical perspectivalism. 

 \section{Conclusion}

There are indeed good scientific reasons, coming both from quantum mechanics and relativity, to think that physical perspectives play a central role in defining our experience of reality. But the scientific evidence appears to favour moderate physical perspectivalism over strong physical perspectivalism. And I have argued throughout this article that even at a more purely philosophical level   the motivations for strong physical perspectivalism seem dubious. 

In light of this, I have two recommendations. The first is that the literature on perspectivalism in quantum mechanics and relativity would do well to keep in mind the possibility of moderate physical perspectivalism as well as more radical forms of perspectivalism.   The second is that this literature would benefit from simply making clearer distinctions between various different forms of perspectivalism, because work on this subject has a tendency to equivocate between what I have called epistemic perspectivalism and physical perspectivalism, and between what I have called strong and moderate physical perspectivalism.    Explicitly adopting the terminology suggested in this article  could help to clarify disagreements and thus better understand  how current formulations of physical perspectivalism relate to earlier philosophical perspectivalisms.

\end{document}